\documentclass[iop]{emulateapj}
\usepackage{apjfonts}

\newcommand{\MAXI}{MAXI~J1409$-$619}
\newcommand{\IGR}{IGR~J14043$-$6148}
\newcommand{\EGR}{3EG~J1410$-$6147}
\newcommand{\SAX}{{\em BeppoSAX}}

\bibpunct{(}{)}{;}{a}{}{,}

\begin{document}

\title{\SAX\ observations of the X--ray pulsar \MAXI\ in low state:
 discovery of cyclotron resonance features}

\author{Mauro Orlandini\altaffilmark{1},
 Filippo Frontera\altaffilmark{1,2},
 Nicola Masetti\altaffilmark{1},
 Vito Sguera\altaffilmark{1},
 Lara Sidoli\altaffilmark{3}
}
\altaffiltext{1}{INAF/Istituto di Astrofisica Spaziale e Fisica Cosmica (IASF)
 Bologna, via Gobetti 101, 40129 Bologna, Italy}
\altaffiltext{2}{Dipartimento di Fisica, Universit\`a degli Studi di Ferrara,
 via Saragat 1, 44122 Ferrara, Italy}
\altaffiltext{3}{INAF/Istituto di Astrofisica Spaziale e Fisica Cosmica (IASF)
 Milano, via Bassini 15, 20133 Milano, Italy}

\shorttitle{\SAX\ observations of \MAXI: discovery of CRFs}
\shortauthors{Orlandini {\em et~al.}}
\slugcomment{The Astrophysical Journal, 747:, 2012 March 1}

\begin{abstract}
The transient 500~s X--ray pulsar \MAXI\ was discovered by the slit cameras 
aboard {\em MAXI} on October 17, 2010, and soon after accurately localized 
by {\em Swift}. We found that the source position was serendipitously 
observed in 2000 during \SAX\ observations of the Galactic plane. Two 
sources are clearly detected in the MECS: one is consistent with the 
position of \IGR\ and the other one with that of \MAXI.  We report on the 
analysis of this archival \SAX/MECS observation integrated with newly 
analyzed observation from {\em ASCA} and a set of high-energy observations 
obtained from the offset fields of the \SAX/PDS instrument.  For the 
ON-source observation, the 1.8--100~keV spectrum is fit by an absorbed power 
law with a photon index $\Gamma = 0.87_{-0.19}^{+0.29}$, corresponding to 
2--10 and 15--100~keV unabsorbed fluxes of $2.7\times 10^{-12}$ and $4\times 
10^{-11}$ erg~cm$^{-2}$~s$^{-1}$, respectively, and a 2--10 keV luminosity 
of $7\!\times\!10^{34}$ erg~s$^{-1}$ for a 15~kpc distance. For a PDS offset 
field observation, performed about one year later and showing a 15--100~keV 
flux of $7\times 10^{-11}$ erg~cm$^{-2}$~s$^{-1}$, we clearly pinpoint three 
spectral absorption features at 44, 73, and 128~keV, resolved both in the 
spectral fit and in the Crab ratio.  We interpret these not harmonically 
spaced features as due to cyclotron resonances. The fundamental energy of 
44$\pm$3~keV corresponds to a magnetic field strength at the neutron star 
surface of $3.8\times10^{12} (1+z)$~G, where $z$ is the gravitational 
redshift.  We discuss the nature of the source in the light of its possible 
counterpart.
\end{abstract}


\section{Introduction}

On October 17, 2010 \citet{ATel2959} reported on an outburst from a new 
source, named \MAXI, revealed by the Gas Slit Camera \citep[GSC;][]{GSC} of 
the Monitor of All-sky X-ray Image \citep[MAXI;][]{MAXI} experiment aboard 
the International Space Station. The 4--10~keV flux decreased from its 
initial value of $\sim$40~mCrab on October 17 to $\sim$30~mCrab on the 
following day.

On October 20 {\em Swift} \citep{Swift} began a target of opportunity 
observation of the \MAXI\ error circle (0\fdg2 radius). \citet{ATel2962} 
used the X--Ray Telescope \citep[XRT;][]{XRT} to pinpoint the position of 
the new source with an estimated 90\% confidence level uncertainty of 
1\farcs9 and coordinates RA(J2000): 14$^{\rm h}$~08$^{\rm m}$~02\fs56, 
Dec(J2000): $-$61\degr~59\arcmin~00\farcs3. Its spectrum is fit by an 
absorbed power law, with photon index $\Gamma=-0.5^{+0.1}_{-0.6}$. A 
follow-up XRT observation on November 30 found the source $\sim$7 times 
brighter than on October 20, with an average 0.3--10 keV flux of $7\times 
10^{-10}$ erg~cm$^{-2}$~s$^{-1}$ \citep{ATel3060}. Only in this latter 
observation a $\sim$500~s pulsation was detected, with a 42\% sinusoidal rms 
modulation.  The source was detected since October 18 also in the 15--50~keV 
energy band by the {\em Swift} Burst Alert Telescope \citep[BAT;][]{BAT} at 
the level of $\sim$30~mCrab \citep{Kennea11}.

Neither catalogued radio nor X--ray sources were present in the {\em Swift} 
error circle, while a 2MASS IR star, the likely IR counterpart of the X--ray 
transient, lays 2\farcs1 from the XRT position \citep{ATel2962}.

The source was also observed on October 22 by the Proportional Counter Array 
\citep[PCA;][]{PCA} aboard {\em RXTE} \citep{XTE}.  The X--ray spectrum 
shows a strong 6.5~keV Iron line (E.W.\ 200~eV), and a continuum modeled by 
reflection or partially covering absorption \citep{ATel2969}.  The 2--10~keV 
flux was about $10^{-10}$ erg~cm$^{-2}$~s$^{-1}$, and was strongly variable 
on time-scales of hundreds of seconds. An observation performed on December 
4 found \MAXI\ at a flux level 6--7 times higher than the October 
observation \citep{Atel3070}, and confirmed the 500~s pulse period observed 
by {\em Swift}.

From observations performed on December 2 and 3 by the GLAST Burst Monitor 
\citep[GBM;][]{GBM} aboard the {\em Fermi} satellite, the presence of a 
500~s double peaked pulsating signal was confirmed \citep{ATel3069}.

Following the \MAXI\ localization by {\em Swift,} we searched for \SAX\ 
\citep{SAX} archival observations with the transient position within the 
Narrow Field Instruments (NFIs) Field of View (FoV).  We found that an 
observation performed in the framework of a Galactic plane survey contained 
the \MAXI\ position \citep{ATel2965}.

In this Paper we report on the spectral analysis of this \SAX\ observation, 
integrated with high-energy data obtained from the offset fields of the PDS 
\citep{PDS} instrument. We further present the analysis of a newly analyzed 
{\em ASCA} observation performed during a Galactic plane survey.

\section{Observations}

\subsection{\SAX/PDS Observations}

We searched our \SAX\ archive for observations covering the \MAXI\ position. 
From our local archive we are able to extract not only products from the NFI 
nominal pointings (also available from ASDC -- the ASI Scientific data 
Center\footnote{\url{http://www.asdc.asi.it/bepposax}}), but also net 
spectra from the PDS offset fields.

\begin{deluxetable*}{ccccccccccc}
\tablecaption{List of PDS pointings covering the \MAXI\ error box
\label{tab:pointings}}
\tablewidth{0pt}
\tablehead{
\colhead{Seq} & \colhead{OPn} & \colhead{Date} & \colhead{Duration} &
\colhead{Source Name} & 
\colhead{Dist} & \colhead{RA} & \colhead{DEC} & \colhead{Type} &
\colhead{PDS rate} & \colhead{S/N} \\
 & & & \colhead{(ks)} & & \colhead{$(\arcmin)$} & \colhead{(J2000)} &
 \colhead{(J2000)} 
 & & \colhead{(15--100 keV c/s)} & \\
}

\startdata
 1 & 08317 & 29/01/2000 &  42 & Gal Plane Sur 2  & 19.6 & 211.3474 & $-$61.88437 &   ON & $0.247\pm0.069$ &  3.6 \\
 2 & 10480 & 07/01/2001 & 220 & Circinus Galaxy  & 27.3 & 212.8952 & $-$61.80250 & MOFF & $0.450\pm0.032$ & 14.1 \\
 3 & 08386 & 07/02/2000 &  52 & Gal Plane Sur 16 & 49.5 & 210.2841 & $-$61.83828 &   ON & $0.086\pm0.049$ &  1.7 \\
 4 & 01652 & 23/02/1997 &  90 & Alpha Cen        & 53.6 & 213.5273 & $-$62.53019 & POFF & $0.086\pm0.052$ &  1.6 \\
 5 & 08484 & 19/02/2000 &  47 & Gal Plane Sur 23 & 70.0 & 213.5936 & $-$61.09281 & POFF & $0.144\pm0.072$ &  2.0 \\
\enddata

\tablecomments{The ``Source Name'' refers, in case of offset fields, to the 
nominal ON pointing.  The distance listed in column 6 refers to the angular 
distance of the PDS pointing from the \MAXI\ position.}

\end{deluxetable*}

Indeed, because of the rocking technique used to derive the PDS background, 
two positions offset by 3\fdg5 with respect to the NFI direction were 
alternatively observed every 96~s.  For fields free of sources, we expect 
that differences between the count spectra in the ``plus OFF'' (thereafter 
POFF) and ``minus OFF'' (thereafter MOFF) positions will be consistent with 
zero. If, on the other hand, a contaminating source is present in one of the 
two offset fields, then the spectral difference corresponds to a net, 
background subtracted PDS spectrum of the offset field.

\begin{figure}
\centering \rotatebox{-90}{\plotone{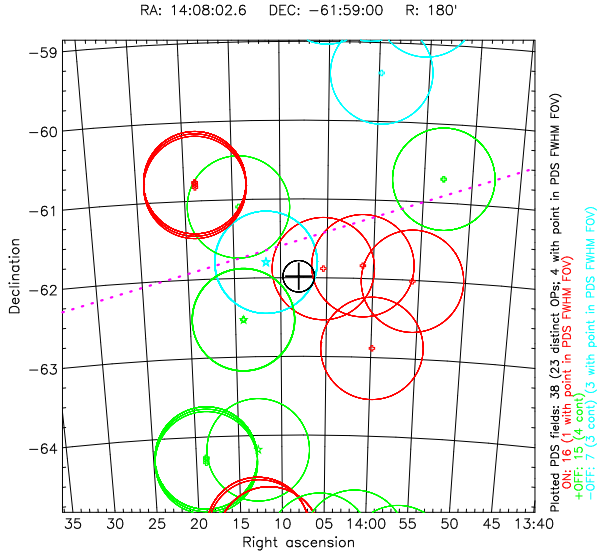}}
\caption{Region centered on the \MAXI\ position (the small black circle) 
with the PDS pointings: ON pointings (red), POFF pointings (green), and MOFF 
pointings (cyan). The PDS FoV is 1\fdg3 (FWHM). Note the two PDS 
observations (one ON and one MOFF) fully covering the source position, and 
three (one ON and two POFF) partially covering the position (with the term 
``partially'' we intend that the source is beyond the collimator FWHM but 
before its FWZI).  The bold dotted line corresponds to the Galactic plane.}
\label{find-chart}
\end{figure}

We found three sets of observations containing the \MAXI\ position in their 
FoVs. In Figure~\ref{find-chart} we show a finding chart centered on the 
\MAXI\ position and showing both the ON-source and OFF-source pointings (the 
PDS FoV is 1\fdg3 FWHM).  In Table~\ref{tab:pointings} we detail the whole 
set of \SAX\ observations used in our analysis. All errors are given at the 
90\% confidence level for a single parameter.

\subsubsection{ON-source observations (OP08317 and OP08386)\label{ON-ana}}

The first ON-source \SAX\ pointing with \MAXI\ in the NFI FoV was performed 
on January 29, 2000 within a program aimed at performing a systematic study 
of part of the Galactic plane (OP08317 -- Galactic Plane Survey Field~02). 
Another ON-source observation (OP08386 -- Galactic Plane Survey Field~16), 
was performed a week later. This observation partially overlaps the OP08317 
field (see Fig.~\ref{find-chart}).

The Galactic plane Field~02 net exposures of the two imaging instruments 
Low-Energy Concentrator Spectrometer \citep[LECS; 0.1--10 keV; 37\arcmin\ 
FoV;][]{LECS} and Medium-Energy Concentrator Spectrometer \citep[MECS; 
1.8--10 keV; 56\arcmin\ FoV;][]{MECS} are 6812 and 22060~s, respectively. 
The difference is due to the constraint that the LECS can be operated only 
when in the Earth shadow. Good data were selected when the instrument 
configurations were nominal, and with an elevation angle above the Earth 
limb $>$4\degr.

\begin{deluxetable*}{cr@{$\,\pm\,$}lcccccccc}
\tablecaption{Sources detected in the MECS field for OP08317
\label{tab:detect}}
\tablewidth{0pt}
\tablehead{
\colhead{\#} & \multicolumn{2}{c}{Rate\tablenotemark{a}} & 
 \multicolumn{2}{c}{pixel} && \colhead{R.A.} & \colhead{Dec} && 
 \colhead{S/N} & \colhead{Counterpart} \\
\cline{4-5} \cline{7-8}
 & \multicolumn{2}{c}{(2--10 keV)} & \colhead{x} & \colhead{y} && 
 \multicolumn{2}{c}{(2000)} \\
}
\startdata
 1 & 45.5 & 3.6  & 373.2 & 317.1 && 14 04 31.3 & $-$61 47 10.8 && 12.6 & \IGR \\
 2 & 10.6 & 1.3  & 188.2 & 233.2 && 14 08 00.4 & $-$61 58 24.1 && 8.15 & \MAXI \\
 3 & 3.77 & 0.79 & 293.9 & 250.1 && 14 06 00.5 & $-$61 56 09.6 && 4.77 & \\
\enddata
\tablenotetext{a}{In units of $10^{-3}$ Counts/s}
\end{deluxetable*}

\begin{figure}
\centering \rotatebox{-90}{\epsscale{1.15}\plotone{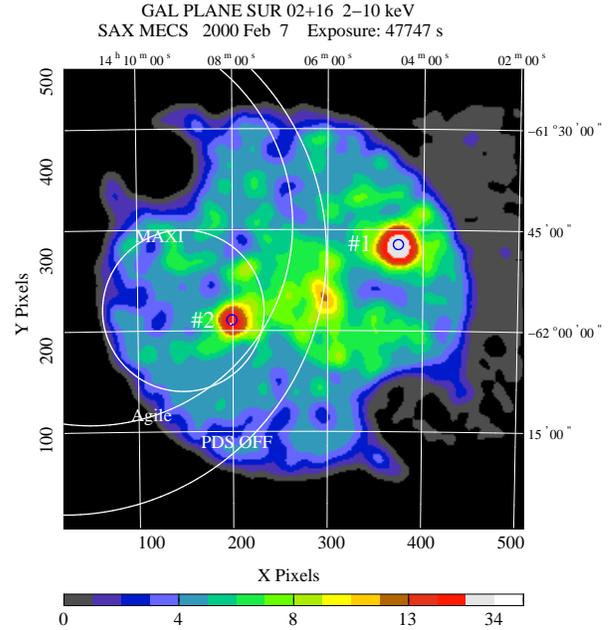}}
\caption{2--10 keV MECS image of a mosaic of the two, partially overlapping, 
ON-source \SAX\ observations OP08317 and OP08386, smoothed with a Gaussian 
filter with a $\sigma$ of 24\arcsec. Two sources are clearly detected. The 
one present in the \MAXI\ error box is listed as \#2 in 
Table~\ref{tab:detect}. The two ``cuts-out'' are due to the removal of the 
calibration source events.}
\label{MECS}
\end{figure}

A mosaic of the two, partially overlapping, ON-source OP08317 and OP08386 
MECS images is shown in Figure~\ref{MECS}, where two relatively bright 
sources are clearly present (see Table~\ref{tab:detect}). Source \#2 
position is consistent with \MAXI, while the source \#1 position is 
consistent with the X--ray source \IGR\ \citep{Bird10}.  The third detection 
(source \#3) corresponds to extended low-energy emission observed in the 
LECS image, and may be associated with a radio source at approximately the 
same position \citep{Cohen01}. The uncertainty in the MECS positions is 
30\arcsec\ \citep{Perri}.

We extracted LECS and MECS data from a 4\arcmin\ (30 pixel) radius circular 
region centered on the \MAXI\ position. Concerning LECS data, there are not 
enough counts to allow a spectral reconstruction.  On the other hand, we are 
able to extract a 1.8--10 keV MECS spectrum, that we rebinned to a minimum 
of 30 counts per bin to allow the use of $\chi^2$ statistics.  We used a 
response matrix appropriate for the off-axis position of the source. Because 
the position of \MAXI\ is close to the Galactic plane, we checked for 
possible contamination from the Galactic ridge emission. To this end, from 
the MECS image we extracted background spectra from two source-free regions, 
offset by about the same angle as the source (to take into account 
vignetting), and with the same extraction radii.  These backgrounds are 
consistent with the mean background used routinely for MECS observations. 
For this reason the mean background was used in the spectral analysis.

Given that the PDS is not an imaging instrument, before performing joint 
MECS/PDS spectral fits it is necessary to address the problem of determining 
whether the high energy emission observed by the PDS is due to \MAXI\ or to 
the {\em INTEGRAL} source. Bear in mind that, in computing fluxes, it is 
necessary to take into account the triangular angular response of the PDS 
collimators \citep[see Fig.~2 in][]{Frontera07}.

First, we analyzed the ON-source OP08386 MECS observation, that contains 
\IGR\ (detected at $>$15$\sigma$) but not \MAXI.  For this observation we 
have only a marginal PDS detection (see Table~\ref{tab:pointings}), 
therefore it seems plausible to associate the high-energy emission observed 
in the OP08317 observation with \MAXI. To further confirm this association, 
we verified that the PDS spectrum extrapolated at lower energies is 
consistent with the \MAXI\ MECS spectrum.

\begin{figure}
\centering \rotatebox{-90}{\epsscale{0.75}\plotone{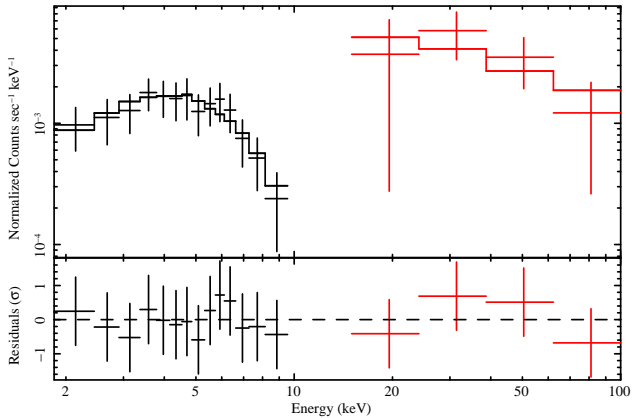}}
\caption{1.8--100 keV joint MECS/PDS count rate spectrum {\em (plus signs)} 
and the power law best fit model {\em (histogram)} of \MAXI\ in the OP08317 
field. The fit residuals are shown in the bottom panel.}
\label{MECS-PDSspectrum}
\end{figure}

In Figure~\ref{MECS-PDSspectrum} we show the joint MECS/PDS 1.8--100 keV 
\MAXI\ count rate spectrum, together with its power law best fit (normalized 
$\chi^2_\nu$ of 0.18 for 15 degrees of freedom -- dof). It is characterized 
by a hard spectrum, with power law photon index $0.87_{-0.19}^{+0.29}$, $N_H 
= 2.8_{-2.2}^{+3.4}\times 10^{22}$ cm$^{-2}$ \citep[consistent with galactic 
absorption,][]{Dickey90} and unabsorbed 2--10 and 15--100~keV fluxes of 
$2.7\times 10^{-12}$ and $4\times 10^{-11}$ erg~cm$^{-2}$~s$^{-1}$, 
respectively.

\subsubsection{MOFF offset field observation (OP10480)\label{MOFF-ana}}

From Figure~\ref{find-chart} we can see that we have three sets of PDS 
offset observations covering the \MAXI\ position. For two of them (OP01652 
and OP08484) we do not have sufficient statistics to perform a useful 
analysis (see the observed count rates in Table~\ref{tab:pointings}).  On 
the other hand, for the MOFF offset field of the Circinus Galaxy 
observation, OP10480, data have enough statistics for a thorough spectral 
analysis.

First, we checked for the presence of any catalogued X--ray sources in the 
MOFF observation FoV.  The only source present is the {\em EGRET} 
\citep{EGRET} source \EGR, that was recently associated with the 50~ms radio 
pulsar PSR~J1410$-$6132 \citep{OBrien08}. No significant X--ray emission has 
been detected around the {\em EGRET} source \citep{Doherty03}; therefore we 
associate the PDS detection with \MAXI.

We started by extracting a 15--100~keV coarser binned PDS spectrum of the 
MOFF field (assuming as its background the corresponding POFF spectrum). The 
fit with a power law gives a photon index of $2.2\pm 0.3$, with a normalized 
$\chi^2_\nu$ of 1.31 for 11 dof. From the shape of the residuals and from 
the quite different photon index with respect to the ON-source observation, 
we suspected the presence of a change of slope in the spectrum between 20 
and 30~keV. This break is not detected in the ON-source observation because 
of its poor statistics. We tried both a broken and a cutoff power law. By 
fixing the cutoff energy at 25~keV (we were only able to constrain the 
break/cutoff energy between 20 and 40~keV), a cutoff power law fit gives a 
power law index $\Gamma=0.9_{-0.3}^{+0.4}$, with a normalized $\chi^2_\nu$ 
of 1.06 for 11 dof. While from a statistical point of view the two fits are 
equivalent (an F-test gives a 36\% probability of chance improvement (PCI) 
of the $\chi^2$ -- see Appendix~\ref{App-cookbook} for the discussion on the 
correctness of the use of the F-test), the cutoff fit nicely matches the 
spectral index observed at lower energies. The corresponding 15--100~keV 
flux, corrected for the triangular response of the collimator, is $7.0\times 
10^{-11}$ erg~cm$^{-2}$~s$^{-1}$.

\begin{figure}
\centering \rotatebox{-90}{\epsscale{0.80}\plotone{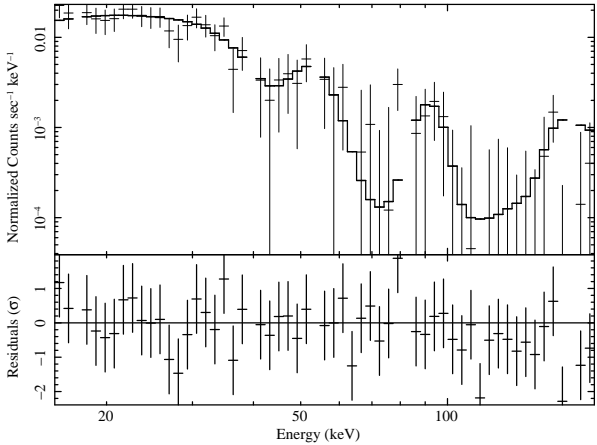}}
\caption{15--200 keV MOFF PDS count rate spectrum {\em (plus signs)} and the 
best fit power law plus 3 Gaussians in absorption model {\em (histogram).} 
The fit residuals are shown in the bottom panel. The Gaussian energies are 
left free and are $44\pm3$, $73_{-5}^{+4}$, and $128_{-8}^{+5}$~keV. The 
line widths were constrained to 4, 4, and 7~keV for the three lines.}
\label{pds_cyc}
\end{figure}

Intrigued by the shape of the residuals, we tried a smaller rebinning factor 
and a wider (15--200~keV) energy range. We were surprised to find the 
spectrum shown in Figure~\ref{pds_cyc}. Absorption features at $\sim$40, 
$\sim$70, and $\sim$120~keV are quite evident. We also noted that the points 
that trail these features at $\sim$50, $\sim$90, and $\sim$150~keV seems to 
be in agreement with a hard continuum up to 200~keV. To test this 
hypothesis, we performed a fit to the data below 30~keV together with these 
trailing points.  No cutoff (the break energy is not constrained) or broken 
power law (the two photon indices are equal) model are able to fit the data, 
while a power law with photon index $1.0\pm0.2$ does ($\chi^2_\nu=0.56$ for 
20 dof).  Therefore the change of slope in 20--40~keV derived from the 
analysis of the coarser binned energy spectrum is really due to the presence 
of the absorption features. The first absorption feature drives the position 
of the break, while the higher energy ones are spread by the rebinning, 
resulting in low-count, high energy bins that are fit by a change of slope.

To further support this interpretation, we performed a joint fit with the 
MECS spectrum (this analysis is detailed in Appendix~\ref{App-fit}).  We 
find that a cutoff or a broken power law models are not able to fit this 
broad band spectrum with the constraint $\Gamma\sim0.9$ below 10~keV.  This 
implies that the origin of the change of slope is not due to the presence of 
a cutoff but to something else, likely absorption features.  For this 
reasons as the continuum we will use a power law with $\Gamma$ fixed at 
$0.87$, as derived by the analysis on the broad band spectrum of the first 
observation.

As throughly discussed in Appendix~\ref{App-cookbook}, the assessment of the 
statistical significance of these absorption features has to be done with 
great care. Our strategy was to evaluate the significance of each feature 
one at a time. We started with a power law fit in the energy range 
15--55~keV, and we obtained an unacceptable $\chi^2_\nu$ of 1.4 for 25 dof. 
The inclusion of a Gaussian in absorption (\texttt{gabs} in \textsc{Xspec}) 
with centroid energy $E_{\rm cyc} = 44\pm3$~keV, $\tau=16_{-7}^{+14}$, and 
width fixed at 4~keV significantly improved the fit ($\chi^2_\nu$ equal to 
0.48 for 23 dof).

Two comments on this measurement are in order: first, the $\sim$44~keV line 
is statistically significant (PCI is $6\times10^{-3}$). Second, the 
difficulties in determining the other line parameters (the line width had to 
be fixed, while the line normalization is affected by large errors) are due 
to the fact that the lines are seen as ``holes'' in the continuum: we have 
not enough statistics to clearly reconstruct the line profile. The line 
width is therefore computed from the ``hole'' width, while for the line 
depth we can infer only an upper limit (this will be more evident for the 
lines at higher energies).

Following our detection strategy, we performed a second fit on the 
15--80~keV energy range. The fit with only one Gaussian in absorption was 
not acceptable: $\chi^2_\nu$ was 3.6 for 32 dof. By adding another Gaussian 
in absorption we obtained a significantly better fit, with $\chi^2_\nu$ of 
0.48 for 30 dof (PCI is $2\times10^{-7}$). The second line parameters are 
$E_{\rm cyc} = 73_{-5}^{+4}$~keV, $\tau=91_{-42}^{+\infty}$, while the the 
line width was fixed at 7~keV. Note how the two line energies are not in an 
harmonic ratio.

Then we extended the energy range to the whole 15--200~keV band. Again the 
fit with two lines was not acceptable ($\chi^2_\nu$ of 4.5 for 46 dof), so 
we added another absorption line, the width of which was fixed at 7~keV. The 
fit significantly improved ($\chi^2_\nu$ of 0.58 for 44 dof, PCI of 
$10^{-10}$), with the best fit $E_{\rm cyc}=128_{-8}^{+5}$~keV. $\tau$ was 
not constrained (this was not a surprise because, as discussed above, we 
have no signal in the ``hole'', and therefore we are not able to put a limit 
to the line normalization).

Finally, we constrained the line energies to be in an harmonic ratio. In 
this case we obtain a $\chi^2_\nu$ of 1.22 for 46 dof. The best fit 
parameters of the fundamental\footnote{Thereafter we will indicate the 
fundamental resonance as (1:1), meaning with this the ordinal number of the 
resonance with respect to the fundamental. With this notation the first 
harmonic will be indicated as (1:2), the third (1:3), and so on.} were 
$E_{\rm cyc}=41\pm1$~keV, $\tau=11_{-7}^{+9}$. All the line widths were 
fixed, as in the previous case with the line energies left free, but we were 
not able to constrain the (1:2) and (1:3) line normalizations.

\begin{figure}
\centering \epsscale{1.15}\plotone{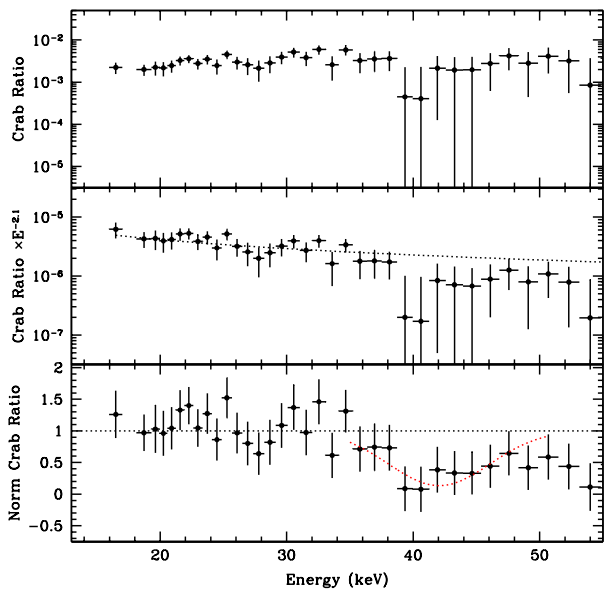}
\caption{{\em Top:} Ratio between \MAXI\ MOFF and Crab count rate spectra. 
{\em Middle:} the \MAXI/Crab ratio multiplied by E$^{-2.1}$, the functional 
form of the Crab spectrum. The dotted line corresponds to the best fit power 
law continuum. {\em Bottom:} ratio between the former expression and the 
best fit power law continuum, together with a Gaussian fit to the 
$\sim$43~keV CRF. The statistical assessment of the CRF has been evaluated 
with a run test, finding that we can reject the line being random at the 
99\% confidence level (see text and Appendix~\ref{App-cookbook} for details).}
\label{crabratio}
\end{figure}

In order to verify that the observed features were not of instrumental 
origin, and at the same time to better characterize the cyclotron resonance 
features (thereafter CRFs), we performed a normalized Crab ratio analysis 
\citep{Orlandini04} on the MOFF spectrum, as we successfully employed for 
the detection of CRFs in numerous X--ray pulsars \citep[see, 
e.g.][]{Orlandini98}. As it is evident from the top panel of 
Figure~\ref{crabratio}, the ratio between the \MAXI\ and the Crab count rate 
spectra shows a ``hole'' in the 38--44~keV range, due to the fundamental 
CRF.  To enhance the feature, we multiplied the \MAXI/Crab ratio by the 
functional form of the Crab spectrum, i.e.\ a simple power law with photon 
index equal to 2.1. The result is shown in the second panel of 
Figure~\ref{crabratio}. Finally, in the lower panel, we show the ratio 
between the previous function and the \MAXI\ best fit continuum power law 
model, together with a Gaussian fit to the fundamental CRF. The Gaussian 
width was fixed at 4~keV, while the Gaussian centroid energy is 
$43_{-1}^{+2}$~keV.  Please note that this value is a lower limit, because 
it does not take into account the intrinsic energy resolution of the PDS 
instrument.

In order to give a quantitative evaluation of the CRF in the normalized Crab 
Ratio we performed a run test (see detailed discussion in 
Appendix~\ref{App-cookbook}). Up to $\sim$34~keV the residuals are 
consistent with random fluctuations ($N_+ = 12$; $N_-=9$; $N_r=13$; 
consistent with random fluctuation at the 84\%). On the other hand, in the 
34--50~keV band we see a clear structure in the residuals. In this case from 
a run test with $N_+=2$, $N_-=14$, $N_r=2$ we we can reject the null 
hypothesis of randomness at the 99\% confidence level.

\subsection{ASCA observation}

The region around \MAXI\ was observed by ASCA on March 2, 1998 for a net 
exposure time of about 18~ks.  The source is clearly detected, and its 
spectrum is confirmed to be quite hard, with a best fit power law photon 
index $\Gamma=0.1_{-0.6}^{+0.8}$ and an unabsorbed 1--10~keV flux of 
$3\times 10^{-12}$ erg~cm$^{-2}$~s$^{-1}$.  The absorption in the direction 
of the source is not well constrained by the fit, with an upper limit of 
$3\times 10^{22}$ cm$^{-2}$, consistent with the {\em Swift} and \SAX/MECS 
results.

\section{A very red(dened) counterpart for \MAXI}

As stressed by \citet{ATel2962}, a NIR source belonging to the 2MASS 
catalogue \citep{Skrutskie06} is found within 2\farcs1 of the \MAXI\ 
position as determined by the XRT. According to that catalogue, this source 
(labeled as 2MASS~J14080271$-$6159020) has NIR magnitudes $J$ = 
15.874$\pm$0.086, $H$ = 13.620$\pm$0.022 and $K$ = 12.560$\pm$0.021. No 
catalogued optical counterpart is present at the position of this NIR 
source, and inspection of the DSS-II-Red digitized archival 
plates\footnote{\url{http://archive.eso.org/dss/dss}} does not show any 
evident optical object at that location. Following \citet{Monet03} we can 
thus place a conservative upper limit, $R\!\!>$20, to the $R$-band magnitude 
of the optical counterpart of this source. The above magnitudes thus 
indicate that this is an extremely red object.

Indeed, assuming the Milky Way extinction law \citep{Cardelli89}, we find 
that the NIR color indices of this source are consistent with those of a 
late O/early B-type star \citep{Wegner94} with a reddening of 
$A_V\!\!\approx \!\!20$ mag. This, using the formula of \citet{Predehel95}, 
implies a column density of N$_{H}$ $\sim$ 3.6$\times$10$^{22}$ cm$^{-2}$, 
which is consistent with that inferred from the {\em Swift} and \SAX\ 
spectral analysis results. This large extinction is also supported by the 
non-detection of the optical counterpart of the source. All this points to a 
likely HMXB nature of this source, hosting a heavily absorbed early-type 
star, similar to several cases of {\em INTEGRAL} hard X--ray sources 
identified as HMXBs at optical and/or NIR wavelengths \citep[see, 
e.g.,][]{Masetti10}.

Assuming then $A_V\!\!\approx\!\!20$ mag along the \MAXI\ line of sight and 
a B0 spectral type for the companion star in this system, we can infer its 
distance, depending on the luminosity class (main sequence, giant or 
supergiant) of the star. Using the tabulated absolute magnitudes for this 
type of star \citep{Lang} we find, for these three cases, respective 
distances of $\sim$4.6, $\sim$7.9 and $\sim$14.5 kpc. Given that the large 
absorption found along the line of sight of this source at both NIR and 
X--ray wavelengths is consistent with the Galactic one, we consider most 
likely that the correct distance is the largest one. Thus, \MAXI\ is quite 
likely located in the far side of the Galaxy, i.e.\ in the farthest parts of 
the Sagittarius-Carina arm, and its mass donor star is likely an early-type 
supergiant.

\section{Discussion}

We found five sets of observations containing the position of \MAXI\ in our 
\SAX/PDS archive, performed in 1997, 2000, and 2001, and one in the {\em 
ASCA} archive (March 1998). In all our observations the source was in a low 
state, with 15--100~keV fluxes in the range $\sim$2--8~mCrab, and no 
spectral variability during the observations. For comparison, an integrated 
exposure (over 5 years) of 2.4~Ms by {\em INTEGRAL\/}/IBIS provides 
2$\sigma$ upper limits on the persistent quiescent emission of 0.2 and 
0.4~mCrab in the 20--40 and 40--100~keV energy bands, respectively 
\citep{ATel2965}. When assuming the source fluxes in outburst as measured by 
{\em Swift} \citep{ATel2962,Kennea11} and {\em RXTE} \citep{Atel3070}, from 
our low state measurement we can infer a dynamic range of 400 in 15--50~keV, 
and of 300 in 2--10~keV.

The discovery of CRFs in the low state spectrum of \MAXI, together with its 
$\sim$500~s pulsations, unambiguously identify the source as an accreting 
X--ray binary pulsar. Only few sources show multiple CRFs, and only two show 
resonances above the (1:2). Three CRFs were observed during the 2004--2005 
outburst of the X--ray pulsar V0332+53 
\citep{Coburn05,Kreykenbohm05,Tsygankov06}, and five (possible six) CRFs 
were discovered in the spectrum of 4U~0115+63 during its 1999 giant outburst 
\citep{Santangelo99,Heindl99,Ferrigno09}. In both cases, deviations from a 
pure harmonic ratio among the CRFs were observed, and were explained in 
terms of departures from a classical dipolar structure of the magnetic field 
in the line-forming region \citep{Nishimura05,Nishimura08}.

The magnetic field strength at the neutron star surface corresponding to 
$E_{\rm cyc}\!=\!44$~keV is $3.8\times10^{12} (1+z)$~G \citep{Canuto77}, 
where $z$, the gravitational redshift, for a typical neutron star of mass 
1.4~M$_\odot$ and radius 10~Km, is about 0.3. When left as free parameters 
in the fit, we found that the best fit CRF line energies do not follow the 
harmonic relation $E_n = n\,E_1$. A slightly non-harmonicity is expected 
when relativistic effects are taken into account \citep[see, 
e.g.,][]{Meszaros}

\begin{equation}
E_n = m_ec^2\,
\frac{\displaystyle\sqrt{1+2n(B/B_{\rm crit})\sin^2\theta} -1}{\sin^2\theta}\,
\frac{1}{1+z}
\label{eq:en-rel}
\end{equation}

\noindent where $m_e$ is the electron rest mass, $c$ the speed of light, 
$\theta$ the angle between the photon and the magnetic field direction, and 
$B_{\rm crit}=4.414\times 10^{13}$~G is the critical magnetic field strength 
where the cyclotron energy equals the electron rest mass.

\begin{figure}
\centering \epsscale{1.2}\plotone{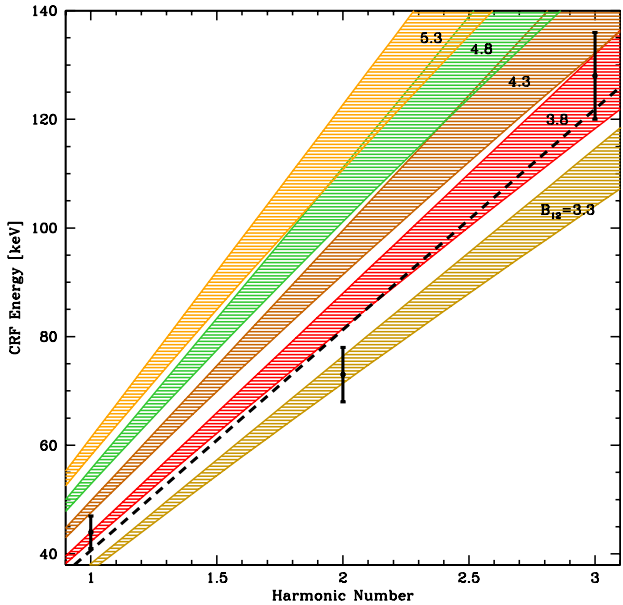}
\caption{Harmonicity of the \MAXI\ CRF line energies. The dashed black line 
corresponds to a linear fit (that is, $E_n=n\,E_1$) to the data, while the 
colored strips take into account relativistic effects, as detailed by 
Eq.~(\ref{eq:en-rel}). Each strip corresponds to a fixed value of the 
magnetic field strength $B_{12}=B\times 10^{12}$~G, and to a range 0.0001--1 
for $\sin^2\theta$, where $\theta$ is the angle between the photon direction 
and that of the magnetic field. While $E_1$ and $E_3$ are consistent with 
$B_{12}\sim3.8$, $E_2$ is consistent with a magnetic field about 20\% lower.}
\label{harmonics}
\end{figure}

The observed (1:2) and (1:3) ratios of the line energies with respect to the 
fundamental, 1.7$\pm$0.2 and 2.9$\pm$0.3, cannot be explained in terms of 
Eq.~(\ref{eq:en-rel}), as it is evident from Figure~\ref{harmonics}. A fit 
to the harmonic relation, shown as the dashed line, gives $E_1=41\pm3$~keV 
(in agreement with the best fit value $E_{\rm cyc}=41\pm1$~keV found when 
imposing the harmonic relation in the spectral fit), but with poor 
significance ($\chi^2_\nu=2.27$ for 2 dof). In the same figure we also show 
the harmonic relation from Eq.~(\ref{eq:en-rel}) for different values of the 
magnetic field and the angle $\theta$.  Taking into account relativistic 
effects, we found that the magnetic field responsible for the (1:1) and 
(1:3) CRFs is about 20\% higher than that responsible for the (1:2) CRF.

Two points are worth noticing: it is always the (1:2) resonance that shows 
the larger disagreement with the harmonic relation, and this could be due to 
the fact that this CRF is due to pure absorption \citep{Nishimura03}, while 
for the other resonances other effects, like multiple scattering and photon 
spawning, enter into play \citep[see, e.g.,][and references 
therein]{Schonherr07}. Unfortunately, because we are not able to reconstruct 
the CRF profiles, we cannot extract more information, like the electron 
temperature and the geometry of the emitting region.

Second, at variance with the V0332+53 and 4U~0115+63 observations, both 
performed during giant outbursts, our \SAX\ observations were performed 
while \MAXI\ was in a low state. This did not allow the study of the 
dependence of the CRF parameters as a function of luminosity, an important 
tool for study of the physical conditions in the line-forming region 
\citep{Mihara04,Nakajima06,Klochkov11}.

The likely early-type optical counterpart and the 500~s pulsation makes 
\MAXI\ a HMXB pulsar. According to the nature of the secondary star we have 
two possibilities: the source is a supergiant fast X--ray transient 
\citep[SFXT;][]{Sguera05,Negueruela06} or a Be/HMXB. In favor of the former 
interpretation is the highly reddened supergiant as possible counterpart, 
the typical outburst X--ray luminosity of $2\times10^{37}$ erg~s$^{-1}$ 
(assuming a distance of 14.5~kpc), and the dynamic range of more than two 
orders of magnitude ($\sim$300) that is typical of the so-called 
``intermediate'' SFXT \citep{Sguera07,Clark10}. Against we have that the 
source active phase, about two months long \citep{ATel3067,Kennea11}, is 
significantly longer that that typical of SFXT \citep[see, 
e.g.,][]{Sidoli09}. If this were the case, then our magnetic field 
measurement would rule out the magnetar nature for SFXTs \citep{Bozzo08}. 
The observed properties of \MAXI\ are also in agreement with those observed 
in other Be/HMXBs, like 1A~1118$-$615, a 400~s X--ray pulsar with a hard 
X--ray spectrum ($\Gamma\!\sim\!1$), a CRF at $\sim$55~keV, and long (tens 
of years) periods of quiescence interrupted by giant (Type-II) outbursts 
lasting weeks to months, in which the X--ray luminosity increases by a 
factor $\sim$200 \citep[see, e.g.,][]{Rutledge07}. This would put the source 
a factor $\sim$2--3 closer.

\begin{figure}
\centering \epsscale{1.15}\plotone{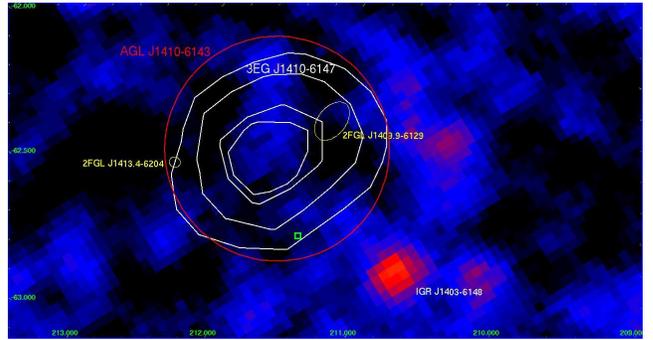}
\caption{Error box of AGL~J1410$-$6143 (big red circle) superimposed on the 
20--100 keV {\em INTEGRAL}/IBIS deep mosaic image ($\sim$2.3 Ms exposure). 
The \SAX/MECS position of \MAXI\ is marked by the green square. The two 
small yellow ellipses represent the Fermi $\gamma$--ray sources 
2FGL~J1409.9$-$6129 and 2FGL~J1413.4$-$6204, respectively. White contours 
(from 50\% to 99\%) refer to the {\em EGRET} source \EGR.}
\label{integral-mosaic}
\end{figure}

Finally, we note that \MAXI\ is located within the $0\fdg5$ error box of the 
unidentified transient MeV source AGL~J1410$-$6143 (see 
Fig.~\ref{integral-mosaic}) discovered on February 21, 2008 by the 
$\gamma$--ray satellite {\em AGILE} \citep{AGILE} during a bright MeV flare 
lasting only about one day \citep{ATel1394,ATel1419}. The {\em AGILE} large 
position uncertainty makes very difficult the identification of its lower 
energy counterpart responsible for the $\gamma$--ray emission. Despite this 
drawback, it is intriguing to note that the flaring source \MAXI\ is the 
only catalogued hard X--ray source above 20 keV (20--100 keV) to be located 
inside the {\em AGILE} error box, according to all the available catalogs in 
the HEASARC database. This spatial correlation is further supported by a 
similar transient nature for both \MAXI\ and AGL~J1410$-$6143.  We also 
point out that this is not a unique case. To date, a few HMXBs have been 
unambiguously detected as flaring MeV sources lasting only a few days 
\citep{Sabatini10,Tavani09,Abdo09}. In addition, there are several other 
HMXBs proposed as best candidate counterparts of unidentified transient MeV 
sources located on the Galactic plane \citep{Sguera09a,Sguera11,Sguera09b}. 
For the sake of completeness, we note that within the large {\em AGILE} 
error box there are a number of high energy MeV sources (see 
Fig.~\ref{integral-mosaic}): i) 2FGL~J1409.9$-$6129 and 2FGL~J1413.4$-$6204 
have been reported in the second Fermi source catalog (Abdo et al. 2011) as 
firmly identified $\gamma$--ray pulsars and this unambiguously excludes 
their association with the transient AGL~J1410$-$6143, ii) \EGR\ is still 
unidentified although it has been likely associated with the $\gamma$--ray 
pulsar 2FGL~J1409.9$-$6129 \citep{OBrien08}.  However, \citet{Wallace00} 
reported a possible MeV flare from \EGR\ lasting a few days on November 
1991. This behavior is at variance with the proposed association with the 
$\gamma$--ray pulsar 2FGL~J1409.9$-$6129 while it is more compatible with 
the flaring nature of AGL~J1410$-$6143.  Further multi-wavelength studies 
(radio, NIR, X--ray and $\gamma$--ray) of the sky region are strongly needed 
and encouraged to shed more light on the nature of such high energy 
emitters.

\section*{Acknowledgements} 

We thank an anonymous referee for helpful comments and suggestions that 
greatly improved the paper. The \SAX\ satellite was a joint Italian-Dutch 
programme. Part of this work is based on archival data, software or online 
services provided by the ASI Scientific Data center (ASDC), and the High 
Energy Astrophysics Science Archive Research Center (HEASARC), provided by 
NASA's Goddard Space Flight Center. This publication makes use of data 
products from the 2MASS archive. The authors acknowledge support from 
ASI/INAF grants I/088/06/0, I/009/10/0 and I/033/10/0 and grant PRIN 
INAF/2009.

\bibliographystyle{aa} 

\appendix

\section{On the statistical significance of absorption features}
\label{App-cookbook}

\twocolumngrid

The analysis of the significance of spectral absorption features, and in 
particular CRFs, has historically been fraught with serious 
misunderstanding, that we attribute to the application of techniques made 
for {\em emission} features to absorption ones.  This has a very profound 
impact on the statistical analysis, as detailed in the following.

First, when searching for spectral features, and especially absorption 
features, the binning of data is crucial: indeed a narrow feature can be 
lost if the data binning is too high \citep[see, e.g.,][page 259]{Eadie}, as 
we showed in the data analysis detailed in Section~\ref{MOFF-ana}. But 
before searching for the best binning factor, it has become customary to 
rebin data in order to achieve a minimum number of counts in each bin. The 
reason of this procedure is that we must avoid bins with few counts if we 
want to use the $\chi^2$ statistics to test for the goodness of our fit (in 
other words, errors must be normally distributed).  A minimum 20~counts per 
bin filter is sufficient to achieve this goal \citep{Cash79}.

The application of this criterion to background subtracted spectra is 
correct for instruments that are not background dominated (that is, they 
have an intrinsically low number of counts per bin), like a detector on the 
focal plane of X--ray optics.  On the other hand, when the net source 
spectrum is obtained by subtracting two high counts spectra this correction 
{\em must not} be applied (instruments like \SAX/PDS or {\em RXTE}/HEXTE are 
background dominated, with a very high number of counts per energy channel). 
Although the net source counts can be very low, they are still Gaussian 
distributed, because resulting from the difference of two normally 
distributed counts.

Second, if features are present in the source spectrum, they (clearly) show 
up in the fit residuals, and the standard goodness of fit estimator, the 
minimization of the $\chi^2$ statistics \citep{Lampton76}, should (clearly) 
indicate that the fit is not ``good'' and an additional component is 
necessary.  The inclusion of a new component to the continuum should result 
in a reduction of the reduced $\chi^2$. In order to assess whether the 
improvement of the $\chi^2$ is due to chance or it is because the new 
component is significant it is customary to use the F-test \citep[see, 
e.g.,][page 160]{Barlow}.

A very important point has to be raised here, because we have two completely 
different approaches whether we are dealing with emission or absorption 
features. In the former case we are dealing, using a \textsc{Xspec} 
terminology, with an {\em additive} component, and we test the null 
hypothesis that the coefficient of the new term is zero \citep[][page 
200]{Bevington}. This is the F-test routine that is incorporated in the 
recent versions of \textsc{Xspec}.

On the other hand, when dealing with an absorption feature, the component is 
{\em not} additive but multiplicative (see routines \texttt{gabs} or 
\texttt{cyclabs}). It is therefore obvious that we cannot use the F-test for 
an additive component, because we cannot vary to zero the new component 
coefficient without modifying the parameters of the original model. The two 
components (the multiplicative model and the original continuum) are 
strongly coupled, and variations in one will affect also the other. 
Therefore we must use a different F-test, as described in \citet[][page 
730]{Press}, that tests the null hypothesis that the observed variances 
comes from the same sample (and this routine is {\em not} incorporated in 
\textsc{Xspec}).  This same test must be used to assess if two different 
models are statistically equivalent or not (and indeed we used it in 
Section~\ref{MOFF-ana} to compare the power law and the cutoff power law 
fits). It is worth noting that the application of the F-test to the 
statistical assessment of CRFs is not affected by the concerns raised by its 
use with emission lines \citep{Protassov02}.

Third, CRFs are very shallow features, with their count rates often at the 
instrument sensitivity level. We therefore expect errors to be dominant, and 
consequently the $\chi^2$ could not be the best estimator of the CRF 
significance.  We need a test that takes into account the structure of the 
line (an information which is lost in the $\chi^2$, because we compute the 
square of the difference between the data and the model). In other words, we 
need to compute what is the probability that a particular structure visible 
in the fit residuals occurs by chance.

\begin{figure}
\centering \rotatebox{-90}{\epsscale{0.7}\plotone{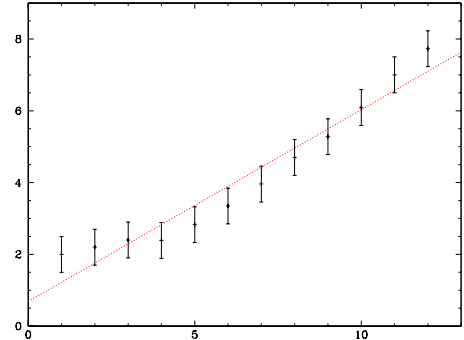}}
\caption{Example on how the goodness of fit estimation performed by the 
$\chi^2$ statistics not always is able to find the best fit model to the 
data. In this case the linear fit to the 12 data points yields a normalized 
$\chi^2$ of 1, but the deviation with respect to the straight line is 
evident. A run test shows that these deviations have only 1.3\%
probability to be due to random fluctuations (see text for details).}
\label{runtest}
\end{figure}

The run test \citep[also known as Wald-Wolfowitz test;][]{Barlow,Eadie} 
works on the {\em signs} of the deviations, that is on the form of the 
residuals. To better clarify how the run test works give a look at 
Figure~\ref{runtest}, adapted from \citet{Barlow}: when fitting the 12 data 
points with a straight line the normalized $\chi^2$ is exactly 1 (likely due 
to error overestimation). But it is evident by eye that the fit is not good 
(indeed the data come from a parabolic model). The reason is that if the fit 
were good we should expect that the number of points ``above'' the fitting 
line should not group together, but should be intermixed with points 
``below'' the fitting line (and this should be more true as the number of 
data points increases). If, on the other hand, we observe only small groups 
of data with the same ``sign'' (called {\em runs\/}), this means that our 
data are not randomly distributed with respect to the fitting model, but 
there is an underlying trend.

As we can see from Figure~\ref{runtest}, we have 12 points, 6 points 
``above'' the fitting line (let us call them $N_+$) and 6 points ``below'' 
the fitting line ($N_-$). The number of runs $N_r$ is only 3, suspiciously 
small. Indeed the probability of obtaining $N_r\le3$ is 1.3\%, telling us 
that the structure observed in the residuals is not due to random 
fluctuations but to a wrong modelization (the linear fit).

From this example, and from the real-case analysis performed on the 
residuals shown in Figures~\ref{crabratio} and \ref{MECS-PDS}, should be 
clear that the goodness of fit in the case of data structures in the 
residuals should not be addressed {\em only} with the $\chi^2$ estimator, 
but it has to be supported by the run test.

In conclusion, this is a sort of {\em cookbook} for a correct evaluation of 
the statistical significance of absorption (CRFs) features in the spectra of 
X--ray sources:

\begin{enumerate}

\item Be careful with the binning: although CRFs are usually broad features, 
a too small binning can hide structures.  When dealing with background 
dominated data (like, for example, \SAX/PDS or {\em RXTE}/HEXTE net spectra) 
{\em never} rebin data in order to have a minimum number of counts per bin, 
otherwise we risk to lose the absorption feature;

\item In order to assess the significance of a CRF the F-test is perfectly 
suitable, but we must use the correct routine: the F-test routine included 
in \textsc{Xspec} is applicable only to an additive component (in 
particular, an emission line), while both \texttt{gabs} and \texttt{cyclabs} 
are multiplicative;

\item The evaluation of the statistical significance of a CRF must be 
supported by other tests, especially in the presence of structures in the 
fit residuals: the run test is able to discriminate whether the structure is 
due to random fluctuations or not.

\end{enumerate}

\onecolumngrid

\section{The continuum spectral model of \MAXI}
\label{App-fit}

\twocolumngrid

The analysis performed on the broad band spectrum of the pointed \MAXI\ 
observation yields as best fit continuum a power law with index $\sim$0.9 
(see section~\ref{ON-ana}).  On the other hand, the analysis performed on a 
coarser binned spectrum of the offset observation indicates the presence of 
a change of slope around 20--40~keV (see section~\ref{MOFF-ana}). Because a 
finer rebinning of this second observation revealed the presence of features 
that can be explained as CRSFs, it is important to understand whether this 
change of slope is due to a cutoff in the continuum or is a combined effect 
of the binning and the presence of the absorption features.

From the pointed observation we know that data below 10~keV must be 
described by a power law with $\Gamma\sim0.9$. To take into account this 
constraint we performed a joint fit of the high energy PDS spectrum from the 
second observation with the low energy MECS spectrum from the first 
observation.

Although the two spectra come from different observations, X--ray pulsars 
are known to not show any spectral variability besides that at low energy 
due to reprocessing of X--rays by the circumstellar material (see, for 
example, the case of Vela X--1 or GX301--2). Because our source does not 
show any intrinsic absorption or emission lines, signatures of reprocessing, 
we are confident that our MECS spectrum did not change between the two 
observations.

\begin{figure*}
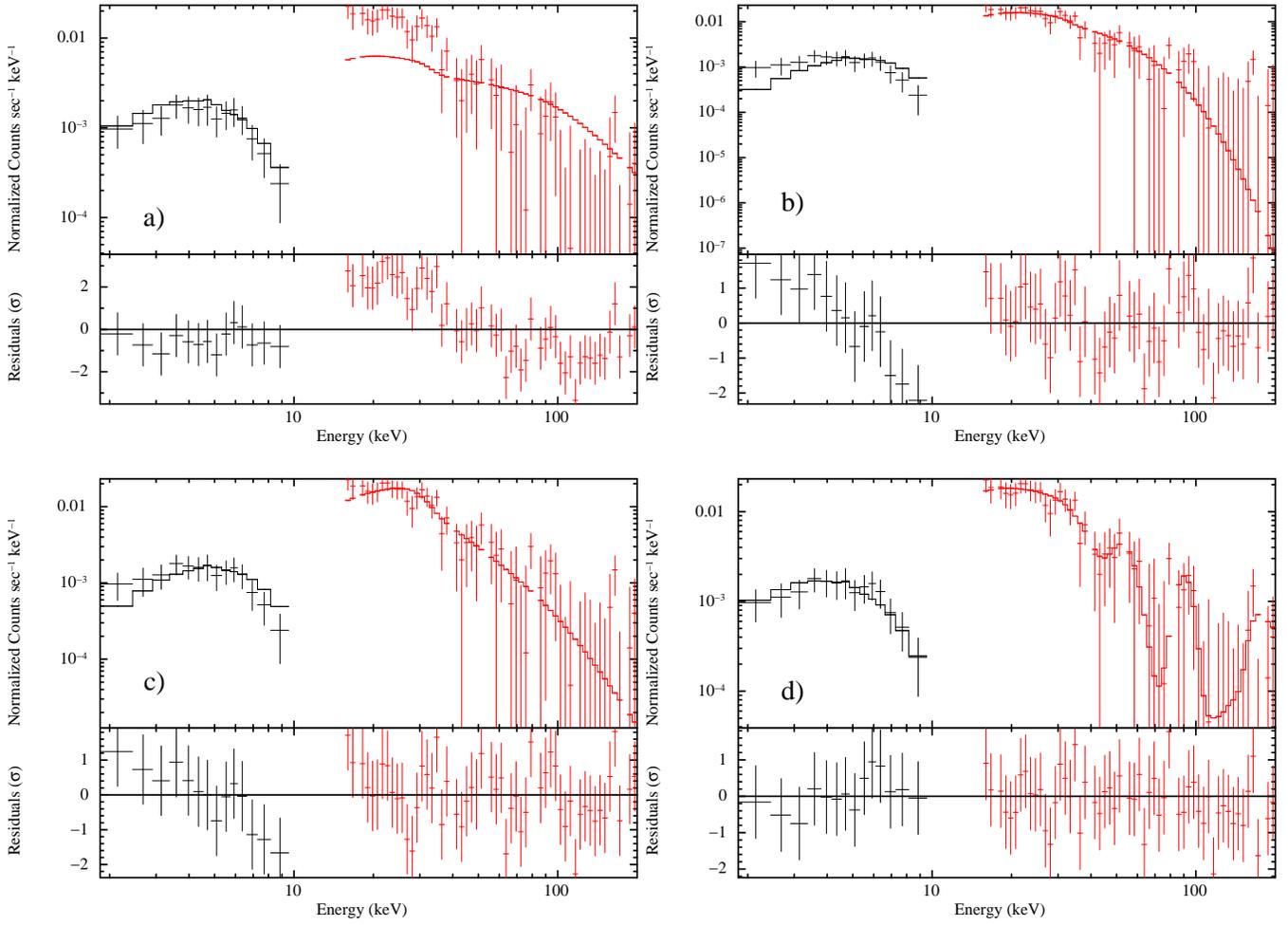

\centering
\includegraphics[width=0.35\textwidth,angle=-90]{fig09a.ps}
\includegraphics[width=0.35\textwidth,angle=-90]{fig09b.ps}

\vspace*{\baselineskip}

\includegraphics[width=0.35\textwidth,angle=-90]{fig09c.ps}
\includegraphics[width=0.35\textwidth,angle=-90]{fig09d.ps}
\caption[]{1.8--200~keV joint MECS (from the pointed observation)~/~PDS 
(from the offset observation) spectrum. {\em a.} Fit with a power law 
continuum. {\em b.} Fit with a cutoff continuum. {\em c.} Fit with a broken 
power law continuum. {\em d.} Fit with a power law continuum and three 
CRSFs.  The only model able to fit simultaneously the low (below 10~keV) and 
high energy data is the power law plus three CRSFs. The best fit parameters 
are listed in Table~\ref{App-table}.}
\label{MECS-PDS}
\end{figure*}

First we tried a fit with a power law, modified at low energy by 
photoelectric absorption.  All the parameters are left free but the $N_H$ 
(fixed at the $2.8\times10^{22}$~cm$^{-2}$ value obtained from the ON-source 
observation), and the result is shown in panel {\em a.}\ of 
Figure~\ref{MECS-PDS}.  The fit parameters are listed in 
Table~\ref{App-table}.

The fit is not statistically acceptable, and the slope is driven by the low 
energy part of the spectrum. To fit also the PDS data we need something that 
is able to change the slope at about 40~keV.

Now let us test whether we can explain this change of slope in terms of a 
cutoff in the continuum. We first tried a cutoff power law (see panel {\em 
b.}\ in Figure~\ref{MECS-PDS}): the fit is statistically acceptable but the 
power law index does not match with that of the low energy part, below 
10~keV. The systematic deviations from the model (information lost by the 
$\chi^2$ test, as discussed in Appendix~\ref{App-cookbook}) can be 
quantified by means of the run test on the MECS residuals ($N_+=8$, $N_-=6$, 
$N_r=3$). The probability of obtaining $N_r\le3$ is 0.5\%, therefore the 
structure observed in the residuals is not due to random fluctuations but to 
the wrong modelization of the continuum. The same result is obtained with 
the broken power law fit (see panel {\em c.}\ in Figure~\ref{MECS-PDS}): 
there is no match to the MECS data.

These results demonstrate that the change of slope is not in the continuum 
but is due to something else: we made the hypothesis that its origin are 
absorption features. Therefore the true continuum model is a power law 
together with CRSF features (see panel {\em d.}\ in Figure~\ref{MECS-PDS}). 
Because we are not able to disentangle the continuum and the CRSFs, due to 
the broadness of the features and their low statistics, we need to fix the 
power law index to the value obtained by the broad band fit.

Just as a test, we performed all the analysis on the CRSFs by fixing the 
power law index at 1.0, 1.1, and 1.2, corresponding to the $1.0\pm0.2$ value 
obtained from the fit to the PDS data below 30~keV and the points trailing 
the CRSFs.  The CRSF parameters are all consistent with each other, 
demonstrating that they do not depend on the particular value of the power 
law index.

\begin{deluxetable}{ccccc}
\tablecaption{Best fit parameters for the joint MECS/PDS spectral fits
\label{App-table}}
\tablewidth{0pt}
\tablehead{ & 
\colhead{Power Law} & \colhead{Cutoff} & 
\colhead{Broken PL} & \colhead{Power Law} \\
& & & & \colhead{+ 3 \texttt{gabs}} \\
}

\startdata
$\Gamma$ & $0.87\pm0.07$ & $-0.75_{-0.5}^{+0.3}$ & 
  $0.06_{-0.2}^{+0.2}$ & $1.0_{-0.2}^{+0.3}$ \\
$E_{\rm cutoff}$ & \nodata & $14\pm4$ & $27_{-4}^{+7}$ & \nodata \\
$\Gamma_2$ & \nodata & \nodata & $2.93_{-0.6}^{+1.9}$ & \nodata \\
$\chi^2_\nu$ (dof) & 2.48 (66) & 0.91 (65) & 0.78 (64) & 0.54 (59) \\
\enddata
\end{deluxetable}

\end{document}